\documentclass[12pt,epsfig]{article}
\pdfoutput=1
\usepackage{amsmath,amssymb}
\usepackage{graphicx}
\usepackage[FIGTOPCAP]{subfigure}
\usepackage{authblk}
\def\beq{\begin{equation}}
\def\eeq#1{\label{#1}\end{equation}}
\def\eeqn{\end{equation}}
\def\beqa{\begin{eqnarray}}
\def\eeqa#1{\label{#1}\end{eqnarray}}
\def\eeqan{\end{eqnarray}}

\def\lsim{\mathrel{\rlap{\lower4pt\hbox{\hskip1pt$\sim$}}
    \raise1pt\hbox{$<$}}}                
\def\gsim{\mathrel{\rlap{\lower4pt\hbox{\hskip1pt$\sim$}}
    \raise1pt\hbox{$>$}}}                

\usepackage{xcolor}
\definecolor{red}{rgb}{1.0, 0, 0}

\def\to{\rightarrow}

\def\stacksymbols #1#2#3#4{\def\theguybelow{#2}
    \def\vp{\lower#3pt}
    \def\sp{\baselineskip0pt\lineskip#4pt}
    \mathrel{\mathpalette\intermediary#1}}

\def\intermediary#1#2{\vp\vbox{\sp
     \everycr={}\tabskip0pt
     \halign{$\mathsurround0pt#1\hfil##\hfil$\crcr#2\crcr
              \theguybelow\crcr}}}

\def\gsim{\stacksymbols{>}{\sim}{2.5}{.2}}
\def\lsim{\stacksymbols{<}{\sim}{2.5}{.2}}

\begin{document}

\begin{titlepage}

\begin{center}

{\Huge \bf Neutralino Oscillations at the LHC  }

\vskip.2cm
\end{center}
\vskip1cm

\begin{center}



{\bf Yuval Grossman$^1$, Bibhushan Shakya$^1$, and Yuhsin Tsai$^2$} \\
\end{center}
\vskip 8pt

\begin{center}
	{$^1$ \it Laboratory of Elementary Particle Physics,
	     Cornell University, Ithaca, NY 14853, USA\\
	     $^2$ \it Department of Physics, University of California, Davis, CA 95616, USA
	      } \\

\vspace*{0.3cm}

{\tt  yg73@cornell.edu, bs475@cornell.edu, yhtsai@ucdavis.edu}
\end{center}

\vglue 0.3truecm

\vskip 0.5cm

\abstract
{We study a supersymmetric phenomenon that can give spectacular signals at the LHC: oscillations of neutralinos. Such oscillations can be naturally realised in R-symmetric models, where additional fields are introduced as Dirac mass partners of gauginos and Higgsinos. Majorana masses for gauginos, necessarily generated from anomaly mediation, can create tiny mass splittings between degenerate mass eigenstates, causing states produced at the LHC to oscillate between neutralinos and their Dirac partner fields. Scenarios where such states decay with displaced vertices can lead to striking signatures at the LHC, where the oscillation can be visible directly in the distribution of displaced vertex lengths. We elaborate on the theory and LHC phenomenology of this feature within a specific scenario of a Higgsino decaying with a displaced vertex into a gravitino and a $Z$ boson. 
}

\end{titlepage}

\section{Introduction}

An extensive study of the theoretical and phenomenological aspects of supersymmetry has been one of the most prominent facets of high energy phenomenology for several decades, and these ideas are now finally being probed by the Large Hadron Collider (LHC). However, given the vast richness of theoretical and phenomenological possibilities that supersymmetry allows, it is plausible that signals of new physics can show up in spectacular, unexpected forms in the coming years. While unexpected signals are exciting for phenomenological reasons, they can also be useful in obtaining insights into the details of the underlying models.  With the LHC rapidly providing new data at energy scales never probed before, the study of unexplored phenomenological signatures that could appear at the LHC has become extremely important and timely. 

With this motivation, we study a novel phenomenon that can give spectacular signatures at the LHC but has remained hitherto unexplored: neutralino oscillations. Oscillations within the context of supersymmetry have been discussed in some earlier works; see e.g. \cite{osc,osc3,oscmfvsusy}, where the oscillation of a mesino leaves visible signatures in the content of decay products at a collider. What we have in mind in this work is a more spectacular signal, where the oscillation itself can be directly observed. Such scenarios can be naturally realized with Dirac gauginos in the context of R-symmetric models \cite{rsymmetry,supersoft,manual}, where the traditional Majorana gaugino mass terms are forbidden by R-symmetry, and new partner fields must be introduced to give Dirac masses to the gauginos. However, Majorana masses for gauginos will be generated from supergravity effects via anomaly mediation \cite{outofworld,majoranamass}. If these Majorana masses are small, they introduce tiny mass splittings between degenerate mass eigenstates, enabling a system to oscillate between a gaugino or Higgsino and its partner field. This oscillation behavior can leave visible footprints in collider observables; in particular, if the states in question are long-lived and decay with displaced vertices within the detector, the oscillation can be directly observed in the distribution of lengths of displaced vertices. 

The main purpose of this work is to draw attention to the possibility of this phenomenon and explore whether it can be observed at the LHC. For this reason, we will not discuss in full detail specific model-building questions or perform full-fledged detector level simulations of the signals; such details are premature until hints of such a signal are actually observed. We therefore only pursue these directions at a level sufficient to illustrate that the occurrence and observation of such a striking phenomenon is feasible at the LHC. 

The paper is organized as follows. We begin by discussing the theoretical motivation and details for Dirac gauginos in R-symmetric models in Section \ref{sec:theory}, followed by a discussion of the oscillation phenomenon in Section \ref{sec:oscillation}. In section \ref{sec:model} we elaborate on the collider phenomenology within a specific model of a Higgsino NLSP decaying into a gravitino and a $Z$ boson, and discuss the reconstruction of the oscillation feature at the LHC. Section \ref{sec:others} contains a broader discussion of other scenarios where oscillations can be realized, and some experimental aspects. 

\section{Theoretical Motivation}
\label{sec:theory}


We focus on gauginos and Higgsinos in the context of R-symmetric models -- in particular, the Minimal R-symmetric Supersymmetric Model (MRSSM) \cite{rsymmetry} -- which is well-motivated for several reasons. In this section we discuss how Dirac mass terms arise for the gauginos and Higgsinos in the MRSSM, and Majorana masses for gauginos are necessarily generated via anomaly mediation. We follow the notation in \cite{rsymmetry}. 

The R-symmetry forbids Majorana mass terms for gauginos, requiring the addition of adjoint chiral superfields $\Phi_i$ for each gauge group $i$, with the corresponding fermion carrying the opposite R-charge,  to accommodate R-invariant Dirac masses. The R-symmetry also forbids the $\mu$ term, hence also the Higgsino mass, in the Higgs sector, analogously requiring the introduction of $R_u$ and $R_d$ superfields to partner $H_u$ and $H_d$.  

One can consider the Dirac masses for the gauginos to be elements of a general softly broken supersymmetric theory. If SUSY breaking is assumed to originate from hidden sector spurions, both F and D-type breaking are allowed, which can be written in terms of the spurions as $X=\theta^ 2F$ and $W'_\alpha=\theta_{\alpha} D$. The $W'$ can be thought of as a hidden sector U(1)' that acquires a D-term \cite{dbreaking}. The Dirac gaugino mass originates from the R-symmetric operator involving the D-type spurion \cite{kineticmixing}:
\begin{equation}
\int d^2\theta \frac{W'_\alpha}{M}W^{\alpha}_i\Phi_i\supset m_D\,\tilde{g}_i\,\tilde{g}'_i\,,
\label{gauginomass}
\end{equation}
where M is the mediation scale of SUSY breaking from the hidden sector to the visible sector, $W_i$ represents the gauge superfield, $\tilde{g}$ is the gaugino, and $\tilde{g}'$ is the corresponding Dirac partner. Likewise, in the Higgs sector, the supersymmetric mass term $\mu_uH_uR_u+\mu_dH_dR_d$ arises from 
\begin{equation}
\int d^4\theta\left(\frac{X^\dagger}{M}H_uR_u+\frac{X^\dagger}{M}H_dR_d\right)\,.
\end{equation}

\noindent In addition, one also gets couplings of the electroweak $\Phi_i$ adjoint chiral superfields to the Higgs doublets
\begin{equation}
\label{lambdacouplings}
W_\Phi=\sum_{i=\tilde{B},\tilde{W}} \lambda_u^iH_u\Phi_iR_u + \lambda_d^iR_d\Phi_iH_d\,.
\end{equation}
 These trilinear terms play a vital role, for instance, in electroweak baryogenesis, and large $\lambda_q (q=u,\,d)$ close to the perturbative limit are also favored by a Higgs mass close to $125$ GeV \cite{Fok:2012fb}. These $\lambda_q$ couplings affect the mixing in the neutralino sector and are therefore of relevance.

Despite the R-symmetry, Majorana masses for gauginos cannot be completely eliminated. It is well known that supersymmetry breaking in a general hidden sector model necessarily generates gaugino masses at one-loop as a consequence of the super-Weyl anomaly \cite{outofworld,majoranamass}. This anomaly-mediated Majorana mass term, $m_{\lambda}$, can be obtained through an F-type spurion
\begin{equation}
\displaystyle{\int}\,d^2\theta\,\frac{F}{M_{Pl}} \,W^{\alpha}_i W_{\alpha\,i}\supset m_{\lambda}\,\tilde{g}^{\alpha}\tilde{g}_{\alpha}\,.
\end{equation}
\noindent The consequent Majorana mass for gauginos is given by
\begin{equation}
m_{\lambda}=\frac{\beta(g^2)}{2g^2}m_{3/2}\,,
\end{equation}
where $\beta(g^2)=dg^2/d\ln\mu$ is the gauge beta function 
and $m_{3/2}$ is the gravitino mass. Recall that $m_{3/2}=F/(\sqrt{3}\,M_{Pl})$; hence the sizes of the anomaly-mediated Majorana masses are determined by $\sqrt{F}$, the fundamental scale of SUSY breaking. For a bino, for instance, this corresponds to \cite{onehiggs}
\begin{equation}
m_1=\frac{11\alpha}{4\pi \cos^2(\theta_W)} m_{3/2}\approx8.9\times 10^{-3}m_{3/2}\,.
\label{majoranabino}
\end{equation}
There are no analogous Majorana masses for the R-partners, since the anomaly-mediated mass term is given by the running of anomalous dimension and gauge coupling in the field strength tensor, which does not exist for the R-partners in the superpotential. Likewise, the couplings of the R-partners to Standard Model states are suppressed due to the R-charge assignments, and the only fields they couple significantly to are the corresponding R-scalars. The R-partners therefore have negligible decay widths of their own when these R-scalars are heavier. In the remainder of this paper we assume that this is the case and simply set their widths to zero. 

Therefore, the MRSSM has Dirac masses for gauginos, which requires the introduction of new partner fields. Majorana masses for gauginos are necessarily introduced from anomaly mediation, with their sizes determined by the scale of SUSY breaking. The interplay between the Dirac and Majorana masses can have interesting phenomenological consequences. We have in mind a Dirac mass of $\mathcal{O}(100)$\,GeV and a Majorana mass of  $\mathcal{O}(10^{-5})$\,eV. This hierarchy opens the possibility of neutralino oscillations, to which we now switch our attention.

\section{Oscillations}
\label{sec:oscillation}

To understand the oscillation phenomenon, consider a gaugino $\tilde{g}$ and its adjoint Dirac partner $\tilde{g}' $. In the $(\tilde{g},\tilde{g}' )$ basis, the Hamiltonian takes the form (see for example \cite{cpviolationbook,yuvalreview})
\begin{equation}
{\mathcal{H}}=
\begin{pmatrix}
m_M-i\Gamma/2 & m_D\\
m_D & 0
\end{pmatrix}\,,
\end{equation}
where $m_M$ and $m_D$ are the Majorana and Dirac masses for the gauginos respectively, and $\Gamma$ is the decay width of the gaugino. As discussed in the previous section, we assume that $\tilde{g}' $ has a negligible decay width, and no Majorana mass. Note that this negligible decay width of $\tilde{g}' $ is also responsible for the vanishing of the decay widths in the off-diagonal entries of the Hamiltonian. 

The eigenvalues of this Hamiltonian are
\begin{equation}
\mu_{1,2}=\frac{1}{2}\left(m_M-\frac{i\Gamma}{2}\pm\sqrt{\left(m_M+\frac{i\Gamma}{2}\right)^2+4m_D^2}\,\right)
\end{equation}
For $m_D\gg\Gamma,\,m_M$, the eigenvalues are approximately
\begin{equation}
\mu_{1,2}\approx m_D\pm\frac{1}{2}m_M-\frac{i\Gamma}{2}\left(1\pm\frac{ m_M}{4\,m_D}\right)\,.
\end{equation}
The two eigenstates are therefore approximately degenerate with a common mass $m_D$, with a small mass splitting $m_M$ between them. In this case, the traditional dimensionless ratios $x$ and $y$ are
\begin{equation}
x\equiv\frac{\Delta m}{\Gamma}=\frac{m_M}{\Gamma},~~~~~y\equiv\frac{\Delta \Gamma}{2\,\Gamma}=\frac{m_M}{8\,m_D}\,.
\end{equation}
Since we are working in the $m_D\gg\Gamma,\,m_M$ regime, we set $y=0$ from here on.   

Next, let $\psi_1$ and $\psi_2$ be the two mass eigenstates, with eigenvalues $\mu_1$ and $\mu_2$ respectively. For infinitesimal mass splitting, one has close to maximal mixing, such that we can write
\begin{equation}
\psi_{1,2}\approx\frac{1}{\sqrt{2}}(\,\tilde{g}\pm\tilde{g}' ),~~~~~\tilde{g},\tilde{g}' \approx\frac{1}{\sqrt{2}}(\,\psi_1\pm\psi_2\,)\,.
\end{equation}

\noindent First, consider the more general case 
\begin{equation}
\psi_a=c_{a1}\,\tilde{g}+c_{a2}\,\tilde{g}' ,
\end{equation}
where $a=1,2$. If, at proper time $t=0$, the system is $\psi=x_1\,\tilde{g}+x_2\,\tilde{g}' $, its evolution over time is given by 
\begin{eqnarray}
\label{osc}
\psi\,(t)&=&\,[\,e^{-i\mu_1t}\,c_{11}\,(x_1\,c_{22}+x_2\,c_{21})-e^{-i\mu_2t}\,c_{12}\,(x_1\,c_{21}+x_2\,c_{22})\,]\,\tilde{g}\nonumber\\
&&+\,[\,e^{-i\mu_1t}\,c_{12}\,(x_1\,c_{22}+x_2\,c_{21})-e^{-i\mu_2t}\,c_{22}\,(x_1\,c_{12}+x_2\,c_{11})\,]\,\tilde{g}'\,. 
\end{eqnarray}
Next, consider an interaction that produces a pure gaugino state at $t=0$; this is represented by $(x_1,x_2)=(1,0)$. As evident from Eq.\,(\ref{osc}), at later times this state evolves into a combination of the gaugino $\tilde{g}$ and its partner $\tilde{g}' $. In particular, at time $t$ the gaugino fraction of the state is
\begin{equation}
\langle \tilde{g}|\psi(t)\rangle=e^{-i\mu_1t}\,c_{11}\,c_{22}-e^{-i\mu_2t}\,c_{12}\,c_{21}\approx\frac{1}{2}\,e^{-i\mu_1t}\,(1+e^{-im_Mt})\,.
\end{equation} 
This shows that the gaugino fraction in the produced state oscillates over time. Since only the gaugino, not the Dirac partner, can decay, the probability of decay of the state $\psi$ is also correlated to the gaugino fraction, and is proportional to
\begin{equation}
\label{oscillation}
|\langle\,\tilde{g}|\psi(t)\rangle|^2=\frac{1}{2}\,e^{-\Gamma t}\,[\,1+\text{cos}\,(\,m_Mt\,)\,] 
\end{equation}
We are therefore led to a scenario where an oscillation is superimposed on the exponential decay traditionally expected of an unstable state. If the state in question travels a measurable distance in the detector before decaying, the oscillation can be observed in spectacular fashion: if a large number of such displaced decays are measured, the distribution of the displaced vertex lengths will follow Eq.\,(\ref{oscillation}), offering a tantalizing collider signature of the phenomenon.



\section{A Specific Scenario: Higgsino Oscillation at the LHC}
\label{sec:model}

We now elaborate on the details of the oscillation behavior and the corresponding signal at the LHC within the framework of a specific scenario.  We choose a setup that has been used in literature to study the reconstruction of displaced vertices in \cite{hnlsp}: the decay of a Higgsino NLSP into a gravitino LSP and a $Z$ boson. We adopt this process into an MRSSM framework together with small anomaly-mediated Majorana masses for the gauginos. The Higgsinos have no Majorana masses of their own but mix with the gauginos in the neutralino sector and inherit the oscillation behavior discussed in the previous section, while the leptonic decay channels of the $Z$ enable clean reconstruction of the positions of the displaced vertices, making this an ideal setup in which to study the phenomenon.  

\subsection{Parameters}

For simplicity, we assume that the lightest neutralino is a pure up-type Higgsino, $\tilde{H}_u$, which can decay into the gravitino LSP and the longitudinal component of the $Z$-boson. The decay length of the Higgsino is given by  \cite{hnlsp, Covi:2009bk} 

\begin{equation}
c\tau\approx \left(\frac{\sqrt{F}}{100\,\rm{TeV}}\right)^4\left(\frac{100\,\rm{GeV}}{m_{\tilde{\chi}^0_1}}\right)^5\left(1-\frac{m_Z^2}{m^2_{\tilde{\chi}^0_1}}\right)^{-4}\times 0.2\,\rm{mm}
\label{eq:decaylength}
\end{equation}
where we have assumed sin$^2\beta\approx 1$. The decay width depends mainly on the SUSY-breaking scale $\sqrt{F}$ and the neutralino mass $m_{\tilde{\chi}^0_1}$.

The mass splitting in this case is slightly involved due to the mixing between the gauginos and the Higgsinos in the neutralino sector. Since there is no Majorana mass for Higgsinos, the mass splitting between the lightest degenerate eigenstates (combinations of $\tilde{H}_u$ and $\tilde{R}_u$) arises through the mixing with the gauginos. The wino mass needs to be above $1$ TeV in order to satisfy precision electroweak limits on the $\rho$ parameter \cite{rsymmetry}. Note that if unification of gaugino masses is assumed, this also drives the gluino mass to several TeV, which leads to squark-initiated SUSY processes at the LHC as discussed in \cite{Kribs:2012gx}. The neutralino sector is then essentially reduced to the $\tilde{B}-\tilde{H}_u$ system (while the wino decouples because it is heavy, $\tilde{H_d}$ can be decoupled assuming negligible mixing with this system), for which the neutralino mass matrix in the $\{\tilde{B},\,\tilde{B}',\,\tilde{H}_u,\,\tilde{R}_u\}$ basis is

\begin{equation}
M_N=
\begin{pmatrix}
m_{M (\tilde{B})} & m_{D (\tilde{B})} & m_Z\,\sin\theta_W\,\sin\beta & 0\\
m_{D (\tilde{B})} & 0 & 0& 0\\
m_Z\,\sin\theta_W\,\sin\beta & 0 & 0 & -\mu_u\\
0 & 0 & -\mu_u& 0\\
\end{pmatrix}\,.
\label{neutralinomatrix}
\end{equation}
Here we have also ignored the $\lambda_{q}$ couplings from Eq.~(\ref{lambdacouplings}) for simplicity. Nonzero values for these couplings do not destroy the oscillation behavior discussed in this section; in particular, corrections from their inclusion are negligible for sufficiently heavy R-scalars. These have been checked explicitly. 

With the bino sufficiently heavier than the Higgsino (necessary for the lightest neutralino to be a pure $H_u$), the mass splitting between the two lightest neutralino eigenstates is given by 
\begin{equation}
\Delta m_{\tilde{H}_u}\approx\left(\frac{m_Z\,\sin\theta_W\,\sin\beta}{m_{D (\tilde{B})}}\right)^2 \left(\frac{m_{\tilde{\chi}^0_1}}{m_{D (\tilde{B})}}\right)^2m_{M (\tilde{B})}\,.
\label{eq:higgsinosplitting}
\end{equation}
Since R-symmetry is preserved in the absence of Majorana masses, the dependence on the bino Majorana mass is understandable. The splitting is additionally suppressed by the off-diagonal bino-Higgsino mixing as well as the hierarchy between the bino and Higgsino mass scales; this additional suppression is desirable since it translates into larger oscillation lengths at the LHC, increasing the likelihood of reconstructing the oscillation feature. Since the lightest eigenstate is made up of $\tilde{H}_u$ and $\tilde{R}_u$, we also have $m_{\tilde{\chi}^0_1}=\mu_u$.


For the oscillation to be observable at the LHC, both the oscillation scale and the neutralino decay length in the collider frame must be around mm to m scale, with the decay length greater than the oscillation length so that one or more oscillations can be reconstructed. From Equations (\ref{majoranabino}),\,(\ref{eq:decaylength}), and (\ref{eq:higgsinosplitting}), we see that these two scales depend on the SUSY breaking scale $\sqrt{F}$ and the neutralino mass $m_{\tilde{\chi}^0_1}$. 

\begin{table}
\begin{center}
\begin{tabular}{c|ccccccc}
	\hline\\[-10pt]
	 &$\overline{\tilde{g}}\tilde{g}'$&$\overline{\tilde{W}}\tilde{W}'$&$\overline{\tilde{B}}\tilde{B}'$&$\overline{\tilde{h}_u}\tilde{h}'_u$&$\tilde{t}^*_{L,R}\tilde{t}_{L,R}$&$\overline{\tilde{G}}\tilde{G}'$ \\[2pt]
	\hline\hline
	$\rm mass$ (TeV) & $3$ & $1$ & $0.5$& $0.12$&$0.6$&$F/M_{\rm{pl}}$ \\[2pt]
	\hline
\end{tabular}
\caption{Mass spectrum for Monte Carlo simulations.}
\label{tab:SLEFT}
\end{center}
\end{table} 

We assume the mass spectrum given in Table \ref{tab:SLEFT} (the scalars are assumed to be sufficiently heavy to be decoupled, and not listed). 
A mass splitting tiny enough to slow down oscillations to the collider scale requires a low scale of SUSY breaking. With 
\begin{equation}
\sqrt{F}\simeq 3\times 10^{2} \,{\rm TeV},
\end{equation} 
the mass splitting between the lightest eigenstates is  
\begin{equation}
\Delta m_{\tilde{H}_u}\approx 10^{-5}\,{\rm eV}.
\label{splittingnumber}
\end{equation}
This corresponds to an oscillation scale of c$\tau\approx 2\pi/{\Delta m_{\tilde{H}_u}}\approx 4$ cm. With these scales of SUSY breaking and Higgsino mass,  the proper lifetime of the neutralino is c$\tau\approx20$ cm. These decay and oscillation scales are therefore ideal for the observation of this phenomenon at the LHC. The above SUSY breaking scale also sets the gravitino mass at the eV scale, making it the lightest supersymmetric partner (LSP), as required for the decay channel we are interested in.  

All simulations, plots, and discussions in the following sections are specific to these parameter values, unless specified otherwise. The approximations and discussions of Equations\,(\ref{neutralinomatrix}),\,(\ref{eq:higgsinosplitting}) are provided for intuition only; in our runs, we diagonalize the full neutralino mass matrix and obtain the eigenvalues numerically.

\subsection{LHC Phenomenology: Vertex and Oscillation Reconstruction}

We focus on event reconstruction at the ATLAS detector; the general idea is the same at CMS. The Higgsino NLSP is promptly produced at the interaction point through the decays of stops. Once produced, it travels a macroscopic distance in the collider frame before decaying with a displaced vertex into a gravitino and a $Z$ boson. We focus on events where the $Z$ boson decays into $e^+e^-$, which allows for clean reconstruction of the displaced vertex using measurements in the ECAL and tracker. The ECAL can measure energy, timing, and pointing information, and this is sufficient to reconstruct the vertex as well as the three-momenta of the electron pair. Details of the reconstruction procedure are described in \cite{hnlsp} and will not be repeated here. The background for such displaced vertex events should be negligible, particularly since the lepton pair will also satisfy a $Z$ mass window cut. 

As mentioned earlier, the distribution of displaced vertex lengths should show the oscillation feature denoted in Eq.\,(\ref{oscillation}). There are three main factors that can potentially distort the oscillation signal and make it irrevocable; we now discuss each of them in turn.

\subsubsection {Interaction with Detector Material}

The Higgsino travels through material in the detector before decaying. Although hard scatterings are unlikely since the Higgsino only interacts weakly, passage through detector material can change the oscillation behavior through coherent forward scattering. This is analogous to the matter effect or MSW effect for neutrinos \cite{neutrinomsw,neutrinomsw2}, and its relevance for neutralinos can be estimated by using results from the neutrino MSW effect.

For neutrinos, interaction with electrons in matter induces an effective potential
\begin{equation}
V=\sqrt{2}\,G_F\,N_e,
\label{potential1}
\end{equation}
where $G_F$ is the Fermi constant and $N_e$ is the electron density. This can be written as \cite{joachimthesis}
\begin{equation}
V= 7.56\cdot10^{-14} \left(\frac{\rho}{g/cm^3}\right)Y_e \,{\rm eV}
\label{potential2}
\end{equation}
where $\rho$ is the density of the material and $Y_e$ is the number of electrons per nucleon. Since neutralinos also interact via the weak force, and the number density of quarks is of the same order as $N_e$, Equations\,(\ref{potential1}),(\ref{potential2}) should provide a reasonable order-of-magnitude estimate of the effective potential for the neutralino. With $\rho=10\,{\rm g/cm^3}$ and $Y_e=1$, one obtains $V\approx 10^{-12}$ eV. This is several orders of magnitude smaller than the mass splitting in Eq.\,(\ref{splittingnumber}). Neutralino interaction with detector material is therefore irrelevant for our study.

It might appear surprising that the MSW effect is negligible for neutralinos in this case, where the mass splitting is only $\Delta m\approx 10^{-5}$\,eV, but can be relevant for neutrinos, where the mass splitting is $\Delta m_\nu^2\approx 10^{-5}$\,eV$^2$. The main point is that for the MSW effect to be negligible, the size of the effective potential should be small compared to the oscillation scale, which is given by the difference between the energy eigenvalues. For neutralinos, this is given by $\Delta m$; for neutrinos, it is given by $\Delta m_\nu^2/E$, where $E$ is the neutrino energy. This difference comes from the neutralino being massive, which leads to a different subleading term in the expansion of the energy eigenvalue. For neutrino energy $E\approx10$ MeV, we have $\Delta m_\nu^2/E\approx10^{-12}$\,eV, which is the same scale as the effective potential V.

\subsubsection {Boost Factors}

The convolution of the neutralino lifetime with the effects of boosting to the reference frame of the detector can potentially wash out the oscillation feature.  Variations in boost factors arise due to convolution with the parton distribution functions (PDFs) for the proton and due to kinematics of the various decay processes. The decay probability in the rest frame of the Higgsino is
\begin{equation}
\mathcal{P}(\tilde{h}\to \tilde{G}Z)\propto e^{-\Gamma\, t_{\rm{0}}}\left[\,1+\cos\, (\Delta m \,t_{\rm{0}})\,\right].
\end{equation}
In terms of the time measured in the lab frame, this is
\begin{equation}\label{eq:oscillation_lab}
\mathcal{P}(\tilde{h}\to \tilde{G}Z)\propto e^{-\Gamma\, t_{\rm{lab}}/\gamma}\left[\,1+\cos\,( \Delta m \,t_{\rm{lab}}/\gamma)\,\right],
\end{equation}
and the uncertainty comes from not knowing the value of $\gamma$ on an event-by-event basis. 

To overcome this hurdle, we make use of the fact that events with larger $\gamma$, corresponding to more boosted Higgsinos, are generally correlated with more energetic leptons. Performing appropriate cuts on the energies of the leptons should therefore pick out events with approximately similar boost factors, thereby restoring the oscillation. We verify the efficiency of this method with Monte Carlo simulations. To simulate this process, we generate parton level events with a Higgsino NLSP decaying into a $Z$ and gravitino in  MADGRAPH5 \cite{madgraph5} for the following process at the 14 TeV LHC:
\begin{equation}
p\,p\to \tilde{t}^*_1\,\tilde{t}_1,\qquad \tilde{t}_1\to t\,\tilde{h},\qquad \tilde{h}\to \tilde{G}\,Z, \qquad Z\to l^+\,l^-
\end{equation}
We implement parton-level cuts of $|\eta|\leq 1.5$, lepton $p_T>20$\,GeV, and an angular separation $\Delta R>0.4$ between leptons for greater lepton identification efficiency. In order to reduce the uncertainty of the angular measurement in the ECAL, we also require each lepton to have $E>100$ GeV. Note that there might be some combinatoric background when both stop decays result in the production of a lepton pair via a $Z$ and the wrong leptons get paired, but this should be negligible given the small branching fraction into this final state and the $Z$ mass window cut that the lepton pair is required to satisfy.

\begin{figure}
\centering
\subfigure[ \quad Lepton pair $200< E<2000$ GeV ]{
          \includegraphics[width=0.44\linewidth]{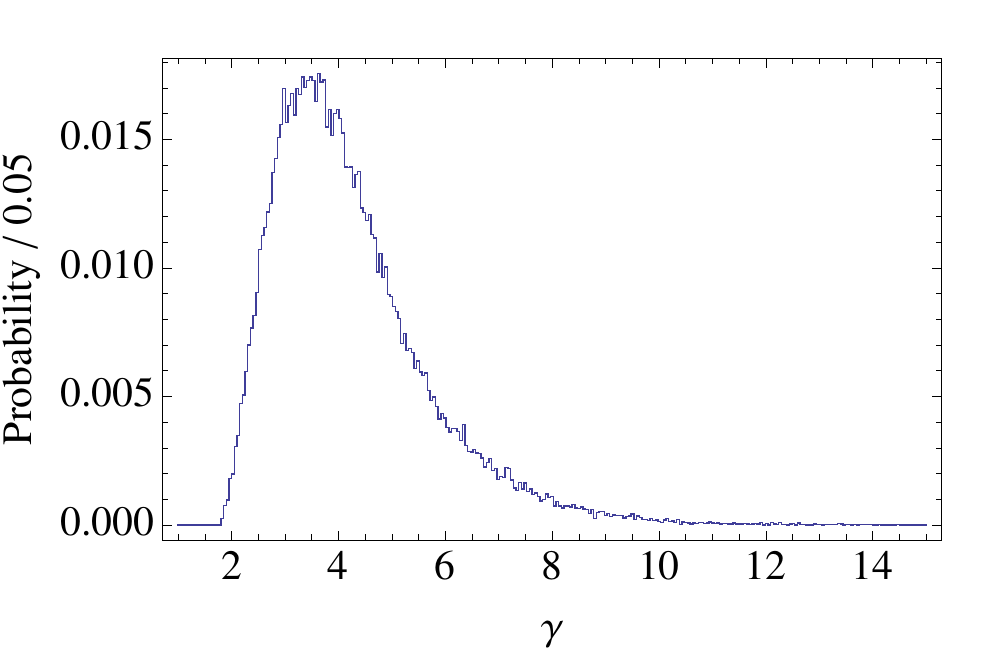}
          \label{fig:gamma400gev}
          }
\subfigure[\quad Lepton pair $200<E<2000$ GeV]{
          \includegraphics[width=0.44\linewidth]{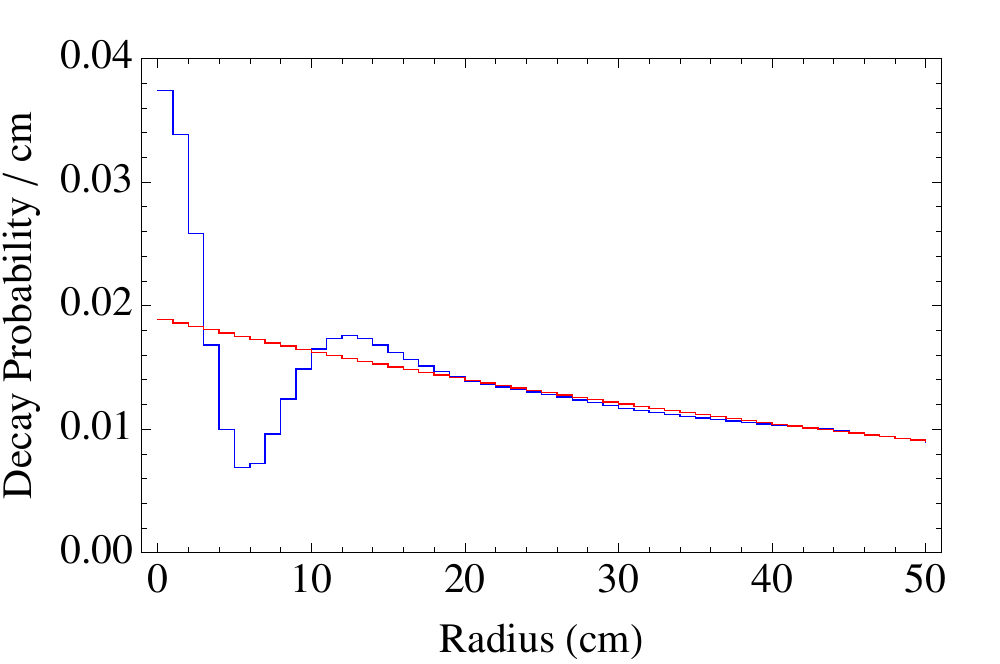}\label{fig:oscillation_plots1}

          \label{fig:oscillat400gev}
          } \\
\subfigure[\quad Lepton pair $500<E<600$ GeV ]{
          \includegraphics[width=0.44\linewidth]{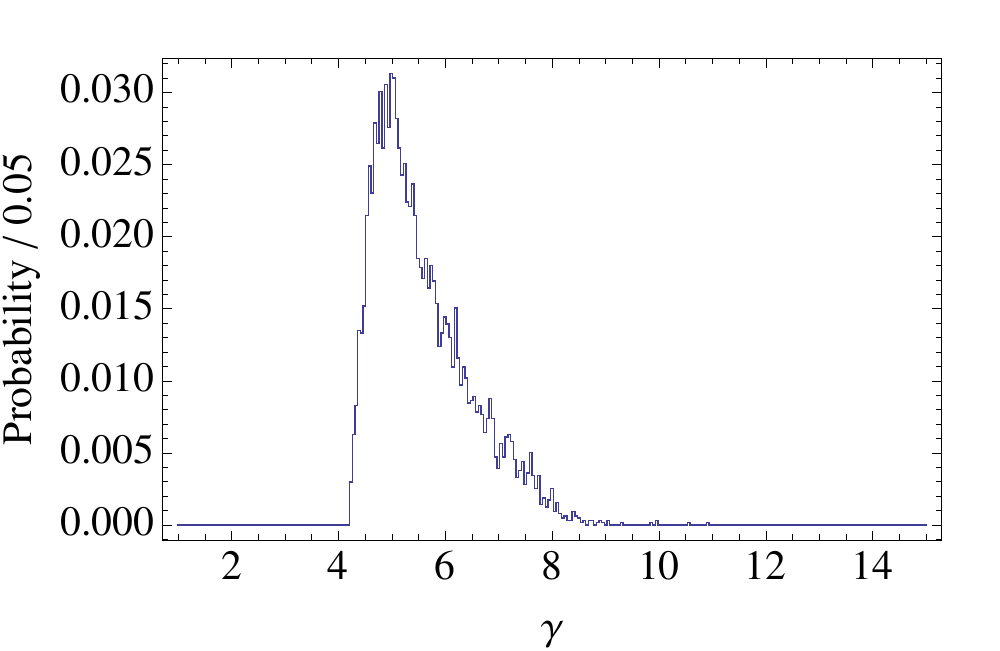}
          }
\subfigure[\quad Lepton pair $500<E<600$ GeV]{
          \includegraphics[width=0.44\linewidth]{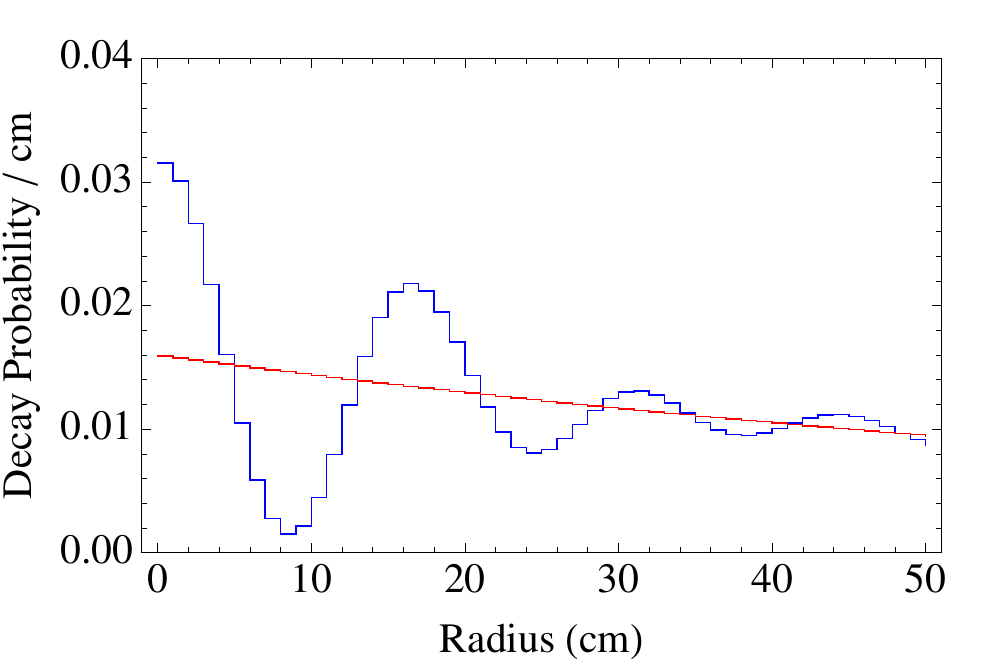}
 } \\ 
\subfigure[\quad Lepton pair $550<E<560$ GeV ]{
          \includegraphics[width=0.44\linewidth]{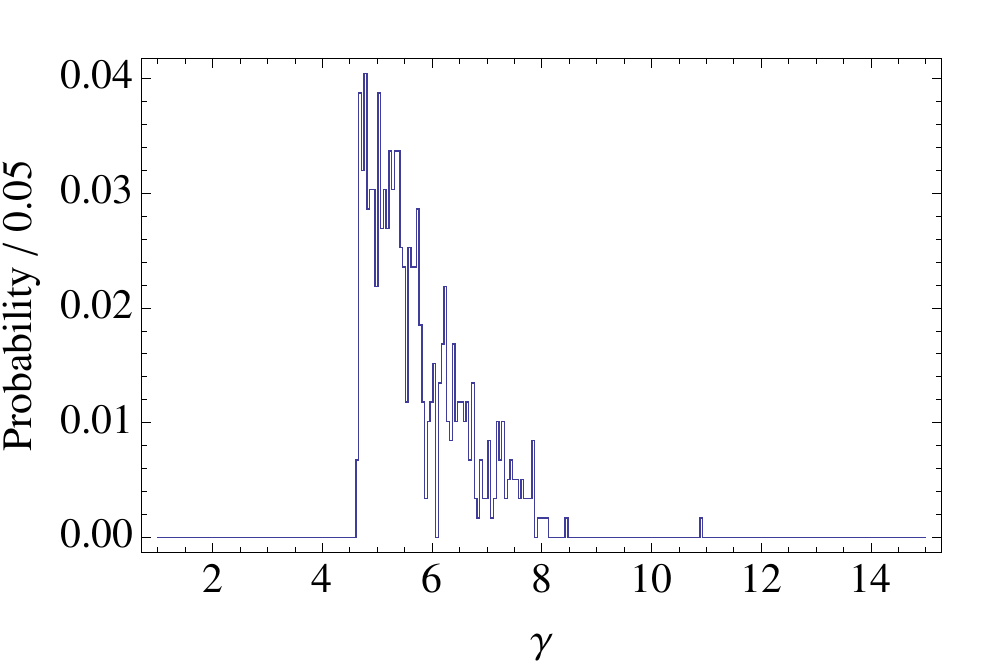}
 }
\subfigure[\quad Lepton pair $550<E<560$ GeV]{
          \includegraphics[width=0.44\linewidth]{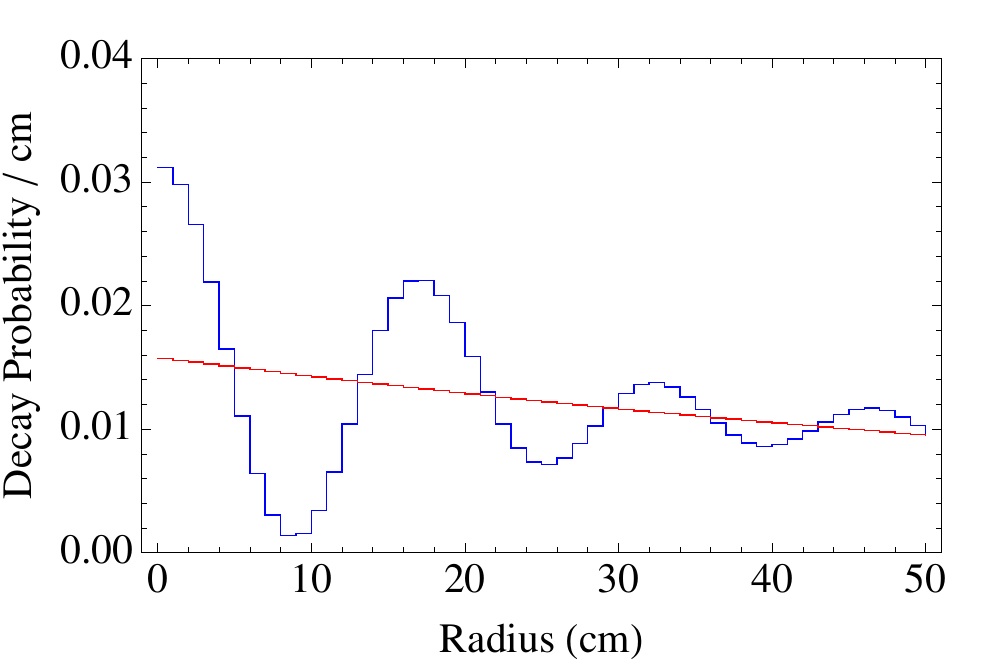}\label{fig:oscillation_plots2}

} \\ 
\caption{Distributions of boost factors and the corresponding distributions of displaced vertex lengths under different lepton energy cuts. The case of decay without oscillation is shown in red. Parton level events are generated using MADGRAPH5 with the cuts and mass spectrum as defined in the text. From top to bottom, the fraction of events that survive the energy cuts are about $100\%$, $11\%$ and $1\%$.}
\label{oscillation_plots}
\end{figure}

The result of performing various cuts on the lepton pair energy is shown in Figure \ref{oscillation_plots}. For various cuts, the left column shows the distribution of boost factors and the right column shows the corresponding distribution of displaced vertices  for the surviving sample (the red line corresponds to what would be observed if the process involved a pure decay with no oscillations). 
It is evident that although the convolution with boost factors obfuscates the oscillation, a narrower cut on the lepton pair energy, which picks out a narrower distribution of $\gamma$, can recover the oscillation feature in the distribution of decay lengths. Note that the energy cut $E>100$ GeV on each lepton also serves as a lower cut on the Higgsino energy, and therefore on the value of $\gamma$. This serves to eliminate oscillation modes with very short oscillation lengths that contribute strongly in washing out the oscillation feature, and is therefore beneficial. It is also worth noting that the Higgsino lifetime in the lab frame is $\mathcal{O}$(ns) (for the $\gamma$ factors in Figure \ref{oscillation_plots}), which will not run into problems with the ATLAS trigger time delay \cite{atlastrigger}.

\begin{figure}
\begin{center}
\includegraphics[width=0.44\linewidth]{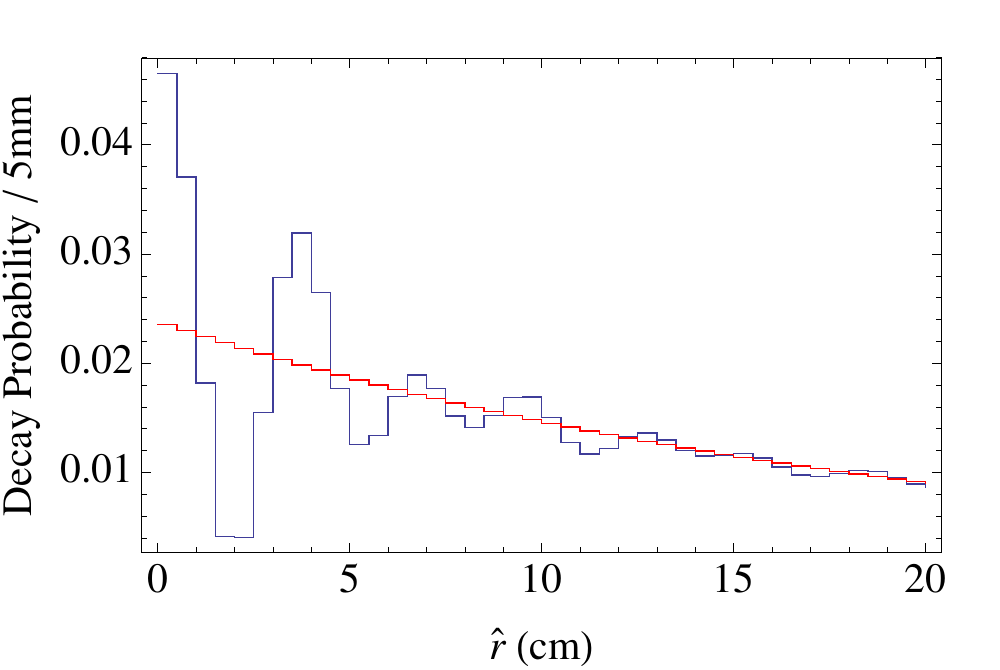}
\includegraphics[width=0.44\linewidth]{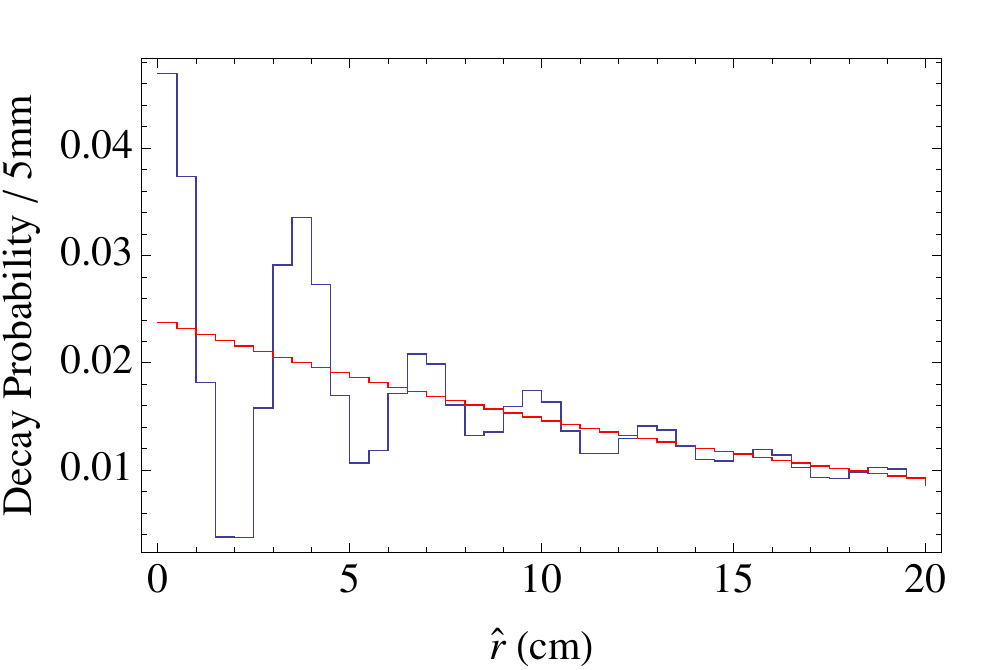}
\end{center}
\caption{Event distribution as a function of $\hat{r}$, for lepton pair energies $200<E<2000$ GeV (left) and $500<E<2000$ GeV (right).}
\label{fig:rhat}
\end{figure}

With a stringent enough cut on the lepton pair energy, the above method is already sufficient to recover the oscillation, albeit at the cost of retaining only a small fraction of the events. More efficient ways of correcting for the boost factor can be implemented if data is scarce. Another approach that uses the correlation between the boost factor and lepton pair energy without throwing away events is scaling down the neutralino decay length by the lepton pair energy, which approximately corrects for the boost factor. For this purpose, define the following quantity to be plotted

\begin{equation}
 \hat{r}\equiv\frac{m_{\tilde{h}}\,r}{E_Z}\,,
\end{equation}
where r is the measured length of the displaced vertex, and $E_Z$ is the energy of the $Z$ boson as reconstructed from the lepton pair energy. Figure \ref{fig:rhat} shows the improvement with using this quantity $\hat{r}$ instead of the measured length of the displaced vertex, where the improvement in reconstructing the oscillation without throwing away a significant number of events is evident. 

\subsubsection {Uncertainty in Vertex Reconstruction}

\begin{figure}
\begin{center}
\includegraphics[width=0.5\linewidth]{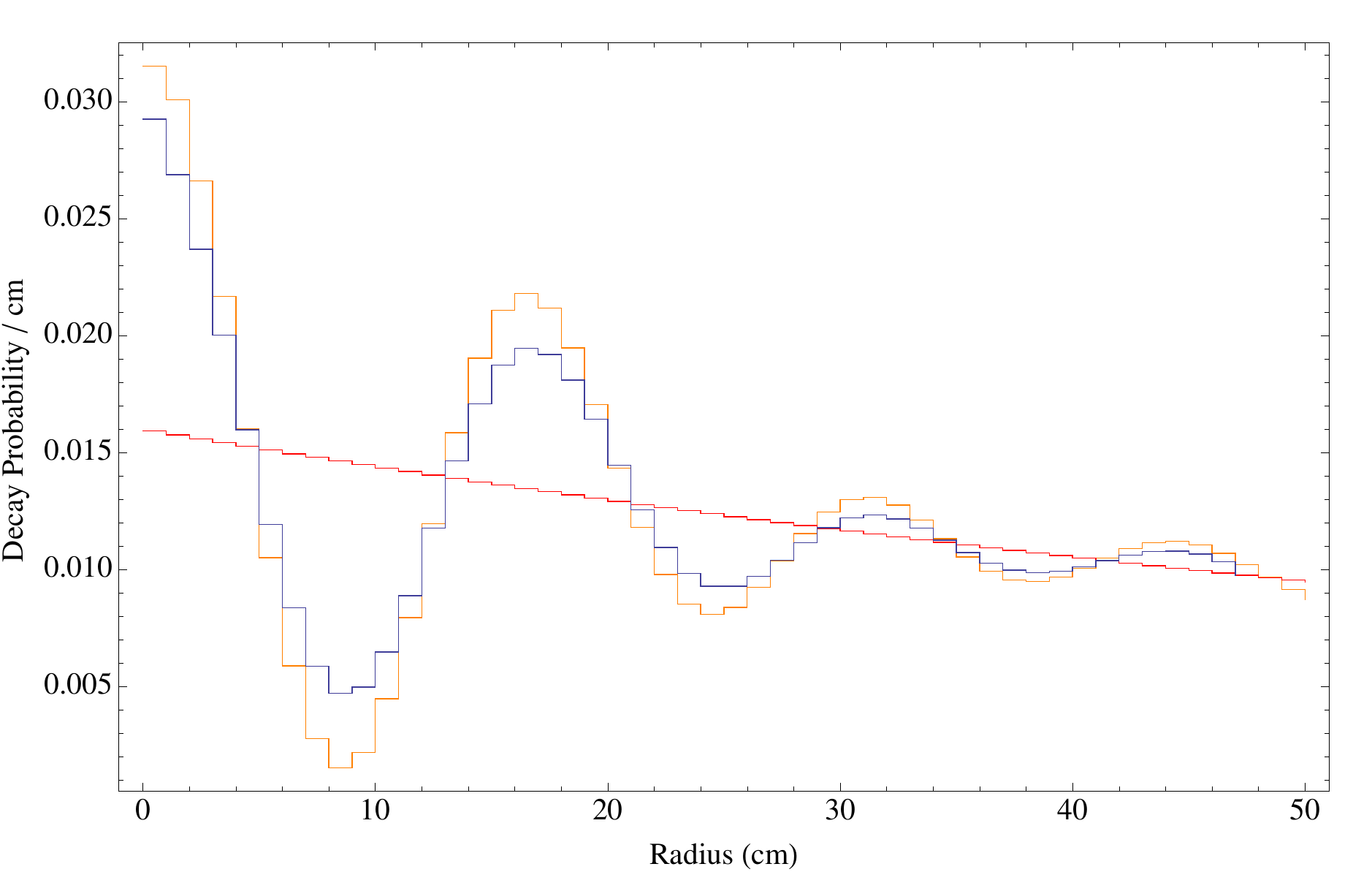}
\end{center}
\caption{The distribution of displaced vertex lengths with pre-selection and lepton energy cut $400<E<500$ GeV. The distribution before (after) smearing by a Gaussian of width $\sigma_r=2$ cm is shown in orange (blue). 
}
\label{fig:smear1}
\end{figure}

There are also uncertainties regarding the reconstruction of the position of the displaced vertex, arising from uncertainties in the measured energies and angles at the detector. We use the uncertainties as listed in \cite{hnlsp}. The largest uncertainty comes from the measurement of $\theta_{dir}$, the angle of the direction of the lepton relative to the beam as measured in the ECAL; it leads to an uncertainty of roughly $\sim2$\,cm in the position of the displaced vertex. We include this uncertainty by smearing the displaced vertex length distribution with a Gaussian of width $2$\,cm (the uncertainty in the reconstruction of the primary interaction vertex is smaller \cite{hnlsp} and can be ignored for this purpose). The distribution of displaced vertex distances before and after the smearing are shown in Figure \ref{fig:smear1}. As we can see, the effect gets slightly washed out but still persists. The oscillation feature can therefore be reconstructed despite measurement uncertainties at the detector. 

\subsection{Estimate of data size and parameter space}

We now estimate the amount of data required to observe the oscillation in the scenario discussed in this section. For the parameters and mass spectrum specified earlier, and with lepton energy cut  $400<E<500$ GeV, we estimate that about 2\% of all $Z\rightarrow l^+l^-$ events are reconstructed and pass all the cuts (this is similar to the efficiency in \cite{hnlsp}). To distinguish the oscillation feature from the case of a pure exponential decay, we require that a fit with the former provide a $3\sigma$ or better fit to the data than a fit with the latter. For this purpose, we only use events with displaced vertex lengths larger than $3$cm, which should have negligible background contamination from prompt events. With these requirements, we estimate the required amount of data to be\begin{equation}  
\mathcal{L}\sim 200\,\rm{fb}^{-1}.
\end{equation}
This corresponds to $\sim 20$ displaced vertex events that pass all the cuts. This particular scenario, at least, is therefore within reach of the LHC within a few years of running. We stress, however, that this is an extremely rough estimate and should be interpreted accordingly, and can change significantly for a different choice of parameters and cuts. 

\begin{figure}
\begin{center}
\includegraphics[width=0.44\linewidth]{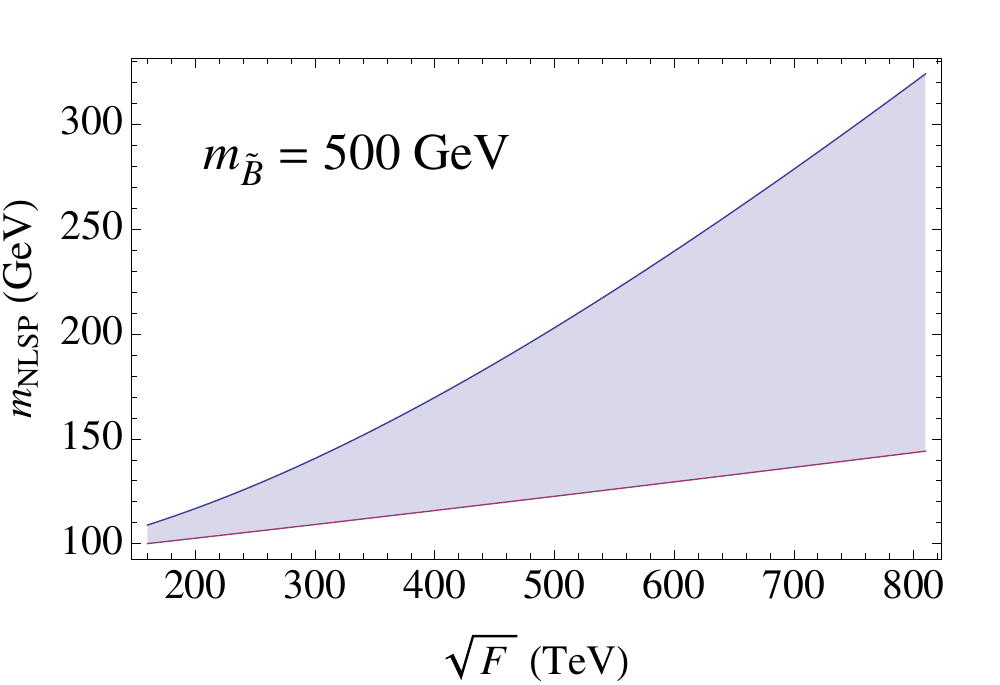},\,\,\includegraphics[width=0.44\linewidth]{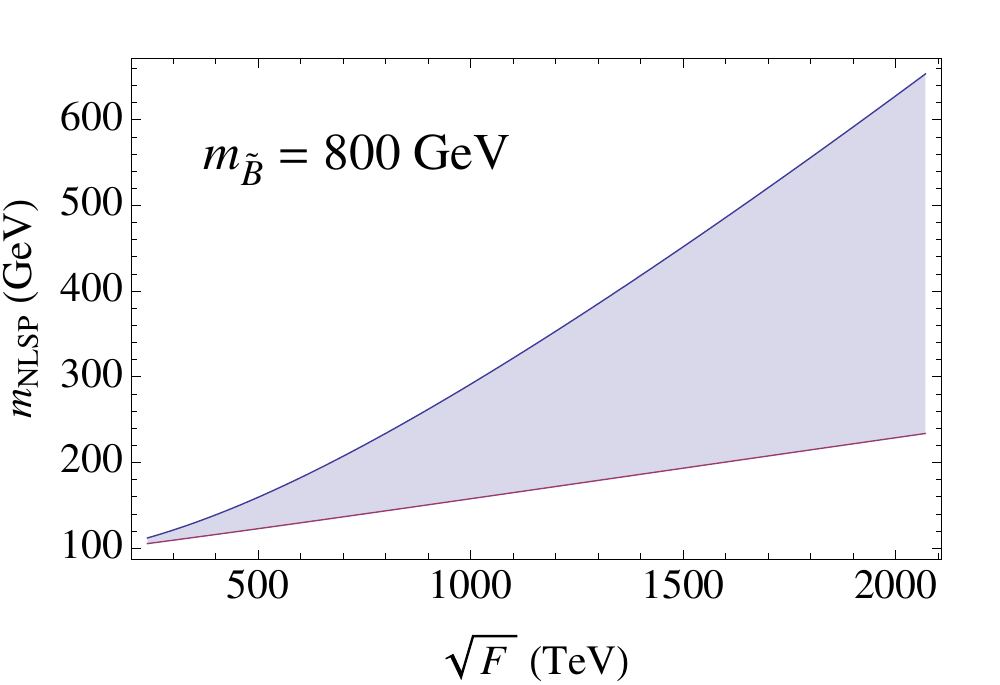}
\end{center}
\caption{Combinations of SUSY breaking scale, $\sqrt{F}$, and higgsino mass, $m_{\tilde{h}}$, for which the oscillation is observable at the LHC with the four constraints discussed in the text, for two different bino masses.}
\label{Fmhbound}
\end{figure}

Next, we explore the regions of parameter space that can give oscillations observable at the LHC. Due to the various experimental constraints discussed in the previous subsection, oscillations cannot be visible for all values of the SUSY breaking scale ($F$) and the higgsino mass ($m_{\tilde{h}}$). In particular, we place the following requirements:
\begin{itemize}
\item The oscillation wavelength ($2\pi\,\Delta m_{\tilde{h}}^{-1}$) must be longer than the precision of the displaced vertex measurement to prevent the feature from being smeared out. Here we require the wavelength to be larger than $2$ cm.
\item The oscillation wavelength must be shorter than $150$ cm in order to see at least one complete oscillation within the ECAL. 
\item The decay length ($c\tau$) must be longer than the oscillation wavelength.
\item The decay length must be shorter than $\sim 4\times 150$ cm, in order to have at least $20\%$ of the decays appear in the measurable region. 
\end{itemize}
Figure \ref{Fmhbound} shows the parameter space that gives oscillations observable at the LHC based on the criteria listed above. We have set $\gamma=6$ to make these plots.The first two constraints set upper and a lower bounds on the SUSY breaking scale $\sqrt{F}$, which are reflected in the limits in the plots. For a given $\sqrt{F}$ and bino mass, the other two constraints set upper and lower bounds on the Higgsino mass. Since the higgsino production rate is not very sensitive to its mass, the observation limit does not depend strongly on $m_{\tilde{h}}$. Improvements in the vertex measurement precision or more stringent energy cuts can easily open up more parameter space. 

\section{Discussions and Conclusions}
\label{sec:others}
Although we have focused exclusively on a single benchmark scenario so far, the oscillation phenomenon is more general and can be realized in many other scenarios. We now turn to a broader discussion of this phenomenon. 

Broadly speaking, the following criteria must be satisfied for the possibility of oscillations at the LHC:
\begin{enumerate}
\item Two approximately degenerate mass eigenstates, separated by a tiny mass splitting, that can be produced in a coherent superposition.  
\item Different interaction behavior (e.g. different decay widths) for the two components that the state oscillates between, in order for the oscillation to be observable. 
\item Decay inside the detector for the signal to be detected.
\end{enumerate}
Any setup that satisfies these requirements can give an interesting oscillation signal at the LHC with the right parameters. 

For instance, Dirac gauginos are not necessary for oscillations in the neutralino sector. The neutralino sector of the MSSM already contains Majorana masses for the bino and neutral wino and Dirac masses for the Higgsinos. If the gauginos are several orders of magnitude heavier than the Higgsinos, the two lightest neutralino mass eigenstates are combinations of $\tilde{H}_u$ and $\tilde{H}_d$, degenerate at Higgsino mass $\mu$, separated by a tiny splitting due to different mixing with the gauginos. The mass splitting in this case is $\mathcal{O}(\mu^2/m_{\tilde{g}})$, which requires the gauginos to be at the GUT scale to get oscillations observable at the LHC. A pure $\tilde{H}_u$ produced in an interaction at the LHC, for instance, will then evolve as a coherent superposition of these two mass eigenstates, oscillating between $\tilde{H}_u$ and $\tilde{H}_d$ as it travels through the detector. If $\tilde{H}_u$ and $\tilde{H}_d$ decay into different Standard Model final states, the oscillation of the neutralino will manifest itself as variations in the final states produced at different distances from the interaction point. 

Likewise, the existence of a gravitino LSP is also not necessary. Neutralinos can decay with detector-scale displaced vertices in models with R-parity violation (see e.g. \cite{displacedsusy}, where R-parity is broken by bilinear terms $\mu_LLH_u$ in the superpotential, and the decays are suppressed due to naturally small coefficients and Yukawa couplings, leading to displaced vertices). If the neutralino sector contains both Dirac and Majorana mass terms, an oscillation signal as described in this paper is possible. 

Moreover, oscillations are not confined to the neutralino sector. In the MRSSM, all gauginos have Dirac masses, and anomaly mediation generates Majorana masses for all gauginos. This implies that the charginos and the gluino can also oscillate into the corresponding Dirac partners. If the lightest chargino (NLSP) is almost degenerate with the lightest neutralino (LSP), the chargino must decay into a three-body final state (e.g. LSP+ virtual $W$, $W$ into two fermions), and can be long lived due to phase space suppression; e.g. \cite{stablecharginos} has postulated charginos with lifetimes $\sim10^{-11}$s. Likewise, the gluino can be long-lived in split-SUSY like scenarios \cite{splitsusy}, where the squarks are heavy and suppress the decay of the gluino. Displaced vertices are therefore possible. The problem with observing oscillations with gluinos and charginos is that they interact via strong and electromagnetic forces.  Long-lived gluinos in split-SUSY are expected to get stopped in the detector. Likewise, ATLAS has looked for long-lived charginos by searching for disappearing tracks in the tracking volume of the detector, where a chargino lighter than 90 GeV with a lifetime between 0.2 and 90\,ns has been excluded \cite{atlaschargino, susysearch}. Unfortunately, such interactions cause the state to decohere into a single mass eigenstate, thereby destroying the oscillation feature. Oscillations with cm-scale displaced vertices extending into the body of the detector are therefore possible only with neutralinos, which interact weakly, but not with gluinos or charginos. 

It can nevertheless be possible to uncover evidence of the oscillation even in the absence of displaced vertices in some scenarios. Consider a scenario where the decay length is comparable to the oscillation length, so that the state completes only a fraction of an oscillation before it decays, but the scale is too small to give rise to displaced vertices, so that the decay is prompt. If the two components that the state oscillates between decay to different Standard Model final states, the time integrated decay fractions into these different states in the presence of an oscillation are different from the fractions that would be expected in its absence. Although not as spectacular as the direct observation of oscillations in the distribution of displaced vertices, observing this deviation in decay fractions would be a hint of an oscillation in effect.  




The observation of oscillations not only provides spectacular signals but can also be useful for extracting valuable information about fundamental parameters in the underlying theory. In the scenario presented in Section\,\ref{sec:model}, for instance, the oscillation scale is tied to the fundamental scale of supersymmetry breaking $\sqrt{F}$; if other parameters of the theory have been measured from other processes, the oscillation signal can be used to calculate the SUSY breaking scale. The observation of the oscillation in the distribution of displaced vertices as discussed in this paper can also be a strong hint of the Dirac nature of gauginos, complementing other signatures of Dirac gauginos and R-symmetric models (see e.g. \cite{collidersignatures,sgluons}).

Improvements on the experimental front can also improve the chances of observing and refining these oscillation signals. The proposed International Linear Collider (ILC) can provide significant improvement, in particular by eliminating the uncertainty from proton PDFs, which is a significant contribution to the smearing of the signal. Improvements in timing techniques at the LHC itself can also significantly improve on measurement uncertainties and corrections for boost factors. 

In summary, in this paper we draw attention to the possibility of a neutralino oscillation signal that can appear at the LHC. Such a signal can be naturally realized in R-symmetric models, where gauginos are Dirac particles but anomaly mediation necessarily generates very small Majorana masses for a low SUSY-breaking scale. The oscillation and decay lengths can be at the right scales for the oscillation to be observable at the LHC. More detailed studies regarding specific model-building issues and detector-level specifics will be required if hints of such a feature actually emerge in the data. For the moment, the possibility of such a spectacular signal at the LHC, which is reasonably well motivated yet has remained hitherto unexplored, is a tantalizing prospect by itself.

\vskip0.8cm
\noindent{\large \bf Acknowledgements} 
\vskip0.3cm

\noindent We especially thank Maxim Perelstein for valuable discussions and comments on the manuscript. We thank Graham Kribs and Matthew Reece for helpful discussions and suggestions. We also acknowledge helpful conversations with Markus Luty, Adam Martin, Patrick Meade, and Rachel Yohay. This research is supported by the U.S. National Science Foundation through grant PHY-0757868. The work of YG is also supported by the United States-Israel Binational Science Foundation (BSF) under grant No. 2010221. YT is also supported by the Department of Energy under grant DE-FG02-91ER406746.



\begin{thebibliography}{99}

  \bibitem{osc} 
  U.~Sarid and S.~D.~Thomas,
  ``Mesino - anti-mesino oscillations,''
  Phys.\ Rev.\ Lett.\  {\bf 85}, 1178 (2000)
  [hep-ph/9909349].
  
  \bibitem{osc3} 
  S.~J.~Gates, Jr. and O.~Lebedev,
  ``Searching for supersymmetry in hadrons,''
  Phys.\ Lett.\ B {\bf 477}, 216 (2000)
  [hep-ph/9912362].
  
\bibitem{oscmfvsusy}
  J.~Berger, C.~Csaki, Y.~Grossman and B.~Heidenreich,
  ``Mesino Oscillation in MFV SUSY,''
  arXiv:1209.4645 [hep-ph].

  \bibitem{rsymmetry}
G.~D.~Kribs, E.~Poppitz and N.~Weiner,
  ``Flavor in supersymmetry with an extended R-symmetry,''
  Phys.\ Rev.\ D\ {\bf 78}, 055010  (2008)
  [arXiv:0712.2039 [hep-ph]].
  
  \bibitem{supersoft}
  P.~J.~Fox, A.~E.~Nelson and N.~Weiner,
  ``Dirac gaugino masses and supersoft supersymmetry breaking,''
  JHEP\ {\bf 0208}, 035  (2002)
  [hep-ph/0206096].
  
  \bibitem{manual}
  K.~Benakli,
  ``Dirac Gauginos: A User Manual,''
  arXiv:1106.1649 [hep-ph].
  
    \bibitem{outofworld}
   L.~Randall and R.~Sundrum,
  ``Out of this world supersymmetry breaking,''
  Nucl.\ Phys.\ B\ {\bf 557}, 79  (1999)
  [hep-th/9810155].
  
  \bibitem{majoranamass}
   G.~F.~Giudice, M.~A.~Luty, H.~Murayama and R.~Rattazzi,
  ``Gaugino mass without singlets,''
  JHEP\ {\bf 9812}, 027  (1998)
  [hep-ph/9810442].
  
      \bibitem{dbreaking}
    Y.~Nomura, D.~Poland and B.~Tweedie,
  ``Minimally fine-tuned supersymmetric standard models with intermediate-scale supersymmetry breaking,''
  Nucl.\ Phys.\ B\ {\bf 745}, 29  (2006)
  [hep-ph/0509243].
  
      \bibitem{kineticmixing}
    K.~Benakli and M.~D.~Goodsell,
  ``Dirac Gauginos and Kinetic Mixing,''
  Nucl.\ Phys.\ B\ {\bf 830}, 315  (2010)
  [arXiv:0909.0017 [hep-ph]].
  
\bibitem{Fok:2012fb} 
  R.~Fok, G.~D.~Kribs, A.~Martin and Y.~Tsai,
  ``Electroweak Baryogenesis in R-symmetric Supersymmetry,''
  arXiv:1208.2784 [hep-ph].
    
   \bibitem{onehiggs}
   R.~Davies, J.~March-Russell and M.~McCullough,
  ``A Supersymmetric One Higgs Doublet Model,''
  JHEP\ {\bf 1104}, 108  (2011)
  [arXiv:1103.1647 [hep-ph]].
  
  \bibitem{cpviolationbook} 
  G.~C.~Branco, L.~Lavoura and J.~P.~Silva,
  ``CP Violation,''
  Int.\ Ser.\ Monogr.\ Phys.\  {\bf 103}, 1 (1999).
  
  \bibitem{yuvalreview} 
  Y.~Grossman,
  ``Introduction to flavor physics,''
  CERN Yellow Report CERN-2010-002, 111-144
  [arXiv:1006.3534 [hep-ph]].
    
\bibitem{hnlsp} 
  P.~Meade, M.~Reece and D.~Shih,
  ``Long-Lived Neutralino NLSPs,''
  JHEP {\bf 1010}, 067 (2010)
  [arXiv:1006.4575 [hep-ph]].
   
  \bibitem{Covi:2009bk} 
  L.~Covi, J.~Hasenkamp, S.~Pokorski and J.~Roberts,
  ``Gravitino Dark Matter and general neutralino NLSP,''
  JHEP {\bf 0911}, 003 (2009)
  [arXiv:0908.3399 [hep-ph]].
    
\bibitem{Kribs:2012gx} 
  G.~D.~Kribs and A.~Martin,
  ``Supersoft Supersymmetry is Super-Safe,''
  Phys.\ Rev.\ D {\bf 85}, 115014 (2012)
  [arXiv:1203.4821 [hep-ph]].

\bibitem{neutrinomsw}
 L.~Wolfenstein,
  ``Neutrino Oscillations in Matter,''
  Phys.\ Rev.\ D {\bf 17}, 2369 (1978).
  
  \bibitem{neutrinomsw2}
   S.~P.~Mikheev and A.~Y.~.Smirnov,
  ``Resonance Amplification of Oscillations in Matter and Spectroscopy of Solar Neutrinos,''
  Sov.\ J.\ Nucl.\ Phys.\  {\bf 42}, 913 (1985)
  [Yad.\ Fiz.\  {\bf 42}, 1441 (1985)].

\bibitem{joachimthesis}
Joachim Kopp, ``Phenomenology of Three-Flavour Neutrino Oscillations." (2006).

\bibitem{madgraph5} 
  J.~Alwall, M.~Herquet, F.~Maltoni, O.~Mattelaer and T.~Stelzer,
  ``MadGraph 5 : Going Beyond,''
  JHEP {\bf 1106}, 128 (2011)
  [arXiv:1106.0522 [hep-ph]].
  
  \bibitem{atlastrigger}
The ATLAS Collaboration,``Triggering on Long-Lived Neutral Particles in the ATLAS Detector," ATLAS Note ATL-PHYS-PUB-2009-082.

  \bibitem{displacedsusy}
    P.~W.~Graham, D.~E.~Kaplan, S.~Rajendran and P.~Saraswat,
  ``Displaced Supersymmetry,''
  arXiv:1204.6038 [hep-ph].
     
  \bibitem{stablecharginos} 
  A.~V.~Gladyshev, D.~I.~Kazakov and M.~G.~Paucar,
  ``Long-lived Charginos in the Focus-point Region of the MSSM Parameter Space,''
  J.\ Phys.\ G G {\bf 36}, 125009 (2009)
  [arXiv:0811.2911 [hep-ph]].
  
    \bibitem{splitsusy} 
  N.~Arkani-Hamed and S.~Dimopoulos,
  ``Supersymmetric unification without low energy supersymmetry and signatures for fine-tuning at the LHC,''
  JHEP {\bf 0506}, 073 (2005)
  [hep-th/0405159].
    
  \bibitem{atlaschargino}
  ATLAS Collaboration, ``Search for long-lived charginos in anomaly-mediated supersymmetry breaking scenarios with the ATLAS detector using 4.7 fb-1 data of pp collisions at sort(s) = 7 TeV", ATLAS-CONF-2012-034

\bibitem{susysearch} 
  S.~Lowette {\it et al.}  [CMS Collaboration],
  ``Supersymmetry Searches with ATLAS and CMS,''
  arXiv:1205.4053 [hep-ex].

  \bibitem{collidersignatures}
  S.~Y.~Choi, D.~Choudhury, A.~Freitas, J.~Kalinowski, J.~M.~Kim and P.~M.~Zerwas,
  ``Dirac Neutralinos and Electroweak Scalar Bosons of N=1/N=2 Hybrid Supersymmetry at Colliders,''
  JHEP\ {\bf 1008}, 025  (2010)
  [arXiv:1005.0818 [hep-ph]].
  
  \bibitem{sgluons} 
  T.~Plehn and T.~M.~P.~Tait,
  ``Seeking Sgluons,''
  J.\ Phys.\ G {\bf 36}, 075001 (2009)
  [arXiv:0810.3919 [hep-ph]].

  \end{thebibliography}

\end{document}